\documentclass[prl,twocolumn,aps,showpacs,amsmath,amssymb]{revtex4-1}

\usepackage{color}
\usepackage{bm}
\usepackage{graphicx}
\usepackage{braket}

\usepackage[plainpages=false,pdfpagelabels,colorlinks=true,linkcolor=red,urlcolor=blue,citecolor=blue,pdftitle={Title},pdfauthor={},pdfdisplaydoctitle=true,pdfduplex=DuplexFlipLongEdge]{hyperref}

\def\Tr{\mbox{Tr}\,}

\newcommand\ba{\begin{eqnarray}}
\newcommand\ea{\end{eqnarray}}
\newcommand\be{\begin{equation}}
\newcommand\ee{\end{equation}}

\begin{document}
\title{Many-body dynamics in long-range hopping model in the presence of correlated and uncorrelated disorder}
\author{Ranjan Modak$^1$ and Tanay Nag$^2$} 
\affiliation{$^1$ SISSA and INFN, via Bonomea 265, 34136 Trieste, Italy}
\affiliation{$^2$ SISSA, via Bonomea 265, 34136 Trieste, Italy}

\begin{abstract}


\textcolor{black}{
Much have been learned about universal properties of entanglement entropy (EE) and participation ration (PR) for Anderson localization. 
We find a new sub-extensive scaling with system size of the above measures 
for algebraic localization as noticed  in one-dimensional long-range hopping models in the presence of uncorrelated disorder.
While the scaling exponent of EE seems to vary  universally with
 the long distance localization exponent  of single particle states (SPSs), PR does not show such university  as it also depends on the
short range correlations of SPSs. On the other hand, in presence of correlated disorder, an admixture of two species of SPSs
(ergodic delocalized and non-ergodic multifractal or localized) are observed, which leads to extensive (sub-extensive) 
scaling of EE (PR). Considering typical many-body eigenstates, we obtain above results  that are further 
corroborated with the asymptotic dynamics.  Additionally, a finite time secondary slow growth in EE is witnessed only for
correlated case while for uncorrelated case there exists only primary growth followd by the saturation. 
We believe that our findings from typical many-body eigenstate would remain unaltered even in the weakly interacting limit.
}

\end{abstract}
\maketitle

\paragraph*{Introduction:}
In one and two dimensions,  an arbitrarily
weak amount of disorder is sufficient to \textcolor{black}{exponentially} localize
all eigenstates  of a system of non-interacting particles, known as Anderson localization 
\cite{anderson.1958,tvr.1979,tvr.1985}. 
However, correlated disorder in one dimensional system can lead to a coexistence of \textcolor{black}{exponentially}
localized and delocalized states, separated by 
mobility edge (ME) \cite{dassarma.1990,ganeshan.2014,modak.2015}. Interestingly,
in the presence of  interactions a transition between delocalized (ergodic) to many-body localized (MBL)
phase  can be observed ~\cite{basko.2006,gornyi2017spectral}. \textcolor{black}{Algebraic localization is another variety of localization that 
draws significant attention in recent times \cite{Burin1989,Levitov1989,Levitov1990,mirlin96,Rodr_guez_2000,Malyshev2004, 
mirlin00,Yao2014,Burin2015,Burin2015b,Gopalakrishnan2017,Nandkishore2017,Tikhonov2018,Luitz2019,de2019algebraic,al_mbl}. 
Entanglement entropy (EE), estimating the bipartite quantum correlations, and participation ratio (PR), 
quantifying  the information about the localization properties {of wave-function}, happen to be the primary measures of localized and 
delocalized phases \cite{huse.2013,serbyn.2013,celardo.2016,vosk.2013}. }



\textcolor{black}{There is an upsurge of studies with disordered models in presence of long range hopping  that decays with
 distance $l$ as a power-law $1/l^a$ can show algebraic localization \cite{singh2017effect,lr1,lr2,nosov2018correlation}.
Interestingly, Levitov's conjecture \cite{Levitov1989,Levitov1990,Burin1989,Malyshev2004} about the absence of 
localization in $d$ dimensional model with $a<d$, is  violated in one of such non-interacting 
long range model where also single particle states (SPSs) are found to be algebraically localized  \cite{deng.2018,Burin1989,nosov2018correlation}.}
Recent advancement in  experiments with 
atomic, molecular and optical systems \cite{exp1,exp2}, power-law
spin interactions with tunable exponent $0 <a < 3$ can
be realized in laser-driven cold atom setup \cite{exp3,exp4}. The 
dipolar ($a=3$) and van-der-Waals ($a=6$)
couplings have also been experimentally observed for the ground-state of neutral
atoms and Rydberg atoms \cite{exp5,exp6,exp7,exp8}.

\textcolor{black}{
Thanks to the availability of analytical and
computational methods, many compelling results have been obtained for EE e.g., it satisfies area (volume) law for exponentially localized 
(delocalized) phase \cite{ferenc.2012,serbyn.2013,agarwal.2014,bardarson.2012,ghosh2019many,bera.2015,huse.2013,sirker.2016}. 
Another important diagnostic PR is expected to follow the similar behavior \cite{celardo.2016,pr.2003}.
} While turning into dynamics, EE for clean (disordered) systems show a faster power law  (a slower double log-type) growth with time 
\cite{Calabrese_2016,calabrese2005evolution, ehud.2013}. 
\textcolor{black}{Having known all of these, we here pose the question that what would be the nature of
EE and PR in an algebraically 
localized phase and do they scale identically? Additionally, the big underlying quest is to predict the Fock space picture for weakly interacting model by performing 
a many-body analysis (statics and dynamics) on the non-interacting system.}


\textcolor{black}{In particular,  we study  EE and PR  for non-interacting  power-law hopping model
in the presence of uncorrelated  disorder (referred as model I), which supports algebraically localized phase,
and correlated disorder (referred as model II), that contains ME and multifractal phases.
We do find that algebraic localization leads to new sub-extensive scaling of EE and PR with system size for model I, while
EE (PR) satisfies  extensive (sub-extensive) scaling for model II. 
Sub-extensive nature of PR is possibly a manifestation of multifractality of many-body wave-functions in Fock space \cite{mirlin06,evers08}. 
Additionally, probing the associated exponents, we can convey that 
EE (PR) can capture the long-range (short-range) correlation of SPSs.
}



\paragraph*{Model:}\label{secII}
We study noninteracting fermions in 1D lattice  in the presence of disordered  potential. The system is 
described by the following  long-range power-law hopping Hamiltonian, 
\begin{eqnarray}
 {H}=-\sum _{i,j\neq i}\frac{1}{|i-j|^a}(\hat{c}^{\dag}_i\hat{c}^{}_{j}+\text{H.c.})+\sum _{i}\epsilon _i \hat{n}_i \nonumber \\
 \label{hamiltonian}
\end{eqnarray}
where $\hat{c}^{\dag}_i$ ($\hat{c}_{i}$) is the fermionic creation (annihilation) operator at site $i$, $\hat{n}_i=\hat{c}^{\dag}_i\hat{c}_{i}$ is the number operator, and $L$ is the size of the system. 

We consider two cases here. 1) Model I (with uncorrelated disorder): $\epsilon_i$ are chosen randomly from a uniform  distribution 
between $[-W,W]$.  
\textcolor{black}{In this paper, we choose $W=20$, for which a very tiny fraction of states are delocalized for $a<3/2$ and all states are localized for $a>3/2$
(see Ref.~\cite{suppl} for details)}. It has been shown that 
the single-particle wave function $\psi(x)$ of this model displays power-law localization ~\cite{Burin1989}
(not exponential) $|\psi(x)|^{2} \sim 1/|x-x_0|^{\nu}$ in the limit $x>>x_0$, where, $x_0$ is the localization center and $\nu$
is the localization exponent. $\nu$ shows
duality $\nu(a)=\nu(2-a)$ around $a=1$ for $0<a<2$ as investigated numerically \cite{deng.2018} \textcolor{black}{ and 
analytically \cite{nosov2018correlation}}. However, we like to point out that near $x_0$, in the limit $|x-x_0|<<L$, the 
SPSs are completely different in both sides of $a$, while one finds an exponential decay
of wave-function 
for $a>1$, 
the decay is still algebraic for $a<1$.  
\textcolor{black}{$a=1$ point has been shown to be critically localized ~\cite{Burin1989}. } 
We note that  
as $a\to \infty$, SPS becomes completely exponentially localized with \textcolor{black}{$L\to \infty$}.
On the other hand, at $a=0$  the model is exactly solvable and wave-functions in the bulk of the spectrum are critically
multifractal ~\cite{Owusu_2008,Ossipov_2013,modak2016integrals,emil.2016}. \textcolor{black}{ We note that the algebraic localization 
is also observed for Hamiltonian  residing in the family of power-law random banded matrix model e.g., model III \cite{Levitov1989,
Levitov1990} (see Ref.~\cite{suppl} for details).}

 2) Model II (with correlated disorder): $\epsilon_i=h\cos(2\pi\sigma i+\phi)$, where 
$\sigma=(\sqrt{5}-1)/2$ and $\phi$ is an offset chosen from a uniform random distribution [0,1]
\textcolor{black}{and closely related to the self-dual quasiperiodic model \cite{Gopalakrishnan2017,Biddle2011}}. 
This seemingly innocent difference has 
drastic consequences on the physics of this  model compared to the previous one. Interestingly, for $a<1$,
all SPSs are extended. However, depending  on the choice of parameters ($h$ and $a$), there are different phases where  ergodic and 
multifractal (MF) states coexist. We will refers this phase as MF phase. 
On the other hand, for $a>1$ there is a coexistence of delocalized and localized SPSs,
hence, mobility edge (ME) exists (we will refer this phase as ME phase). In either side of $a$, different regimes,
denoted by $P_s$, are characterized by a fraction 
\textcolor{black}{$\sigma ^{s}<1$} of ergodic SPSs at the  bottom of the band and the rest are either localized (for $a>1$) or multifractal 
(for $a \leq 1$) (see \cite{suppl} for details) \cite{deng2018one}.
 

For all calculations in this paper, we restrict ourself at half-filling. All quantities are obtained 
after \textcolor{black}{algebraically}  averaging over 
$10^3$ disorder realizations  \textcolor{black}{ 
(see Ref.~\cite{suppl} for details)}.
All time evolution calculations are done 
starting from an initial product state $|\Psi_0\rangle=\prod_{i=1}^{L/2}\hat{c}^{\dag}_{2i}|0\rangle$.

\begin{figure}
\includegraphics[width=0.5\textwidth,height=0.36\textwidth]{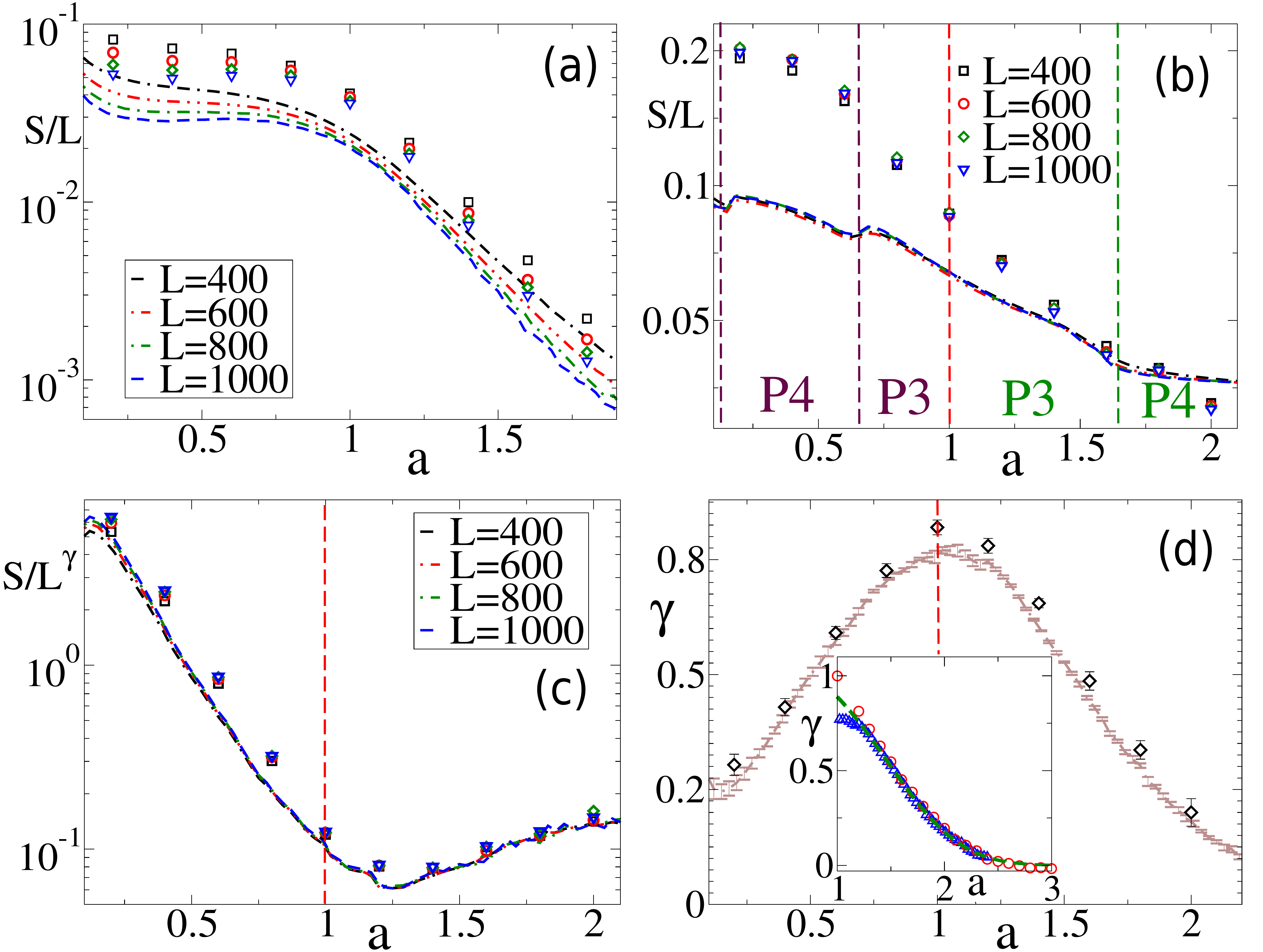}
\caption{(Color online) Entanglement entropy density $S/L$ vs $a$ for  Model I in (a)
and Model II \textcolor{black}{with $h=4.0$} in (b). Here, P3 and P4 represent different multifractal (mobility edge) phases for $a<1$ ($a>1$).
A perfect data collapse is observed in (b) and (c)
for different $L$ suggesting volume law of  EE in model II and sub-extensive scaling of EE in model I
respectively.
 The variation of sub-extensive exponent $\gamma (a)$ with  $a$ showing duality around $a=1$ is depicted in (d). 
Lines (symbols) correspond to  EE obtained from eigenstates (dynamics). \textcolor{black}{Inset in (d) shows that $\gamma$ for model I (depicted by red circle) and 
model in Ref.~\cite{suppl} (depicted by blue triangles) behave identically (referenced with stretched exponential, green dashed line)
with $a$, which is related to localization exponent $\nu\simeq2a$.}
 }
\label{fig3}
\end{figure}

\begin{figure}
\includegraphics[width=0.5\textwidth,height=0.36\textwidth]{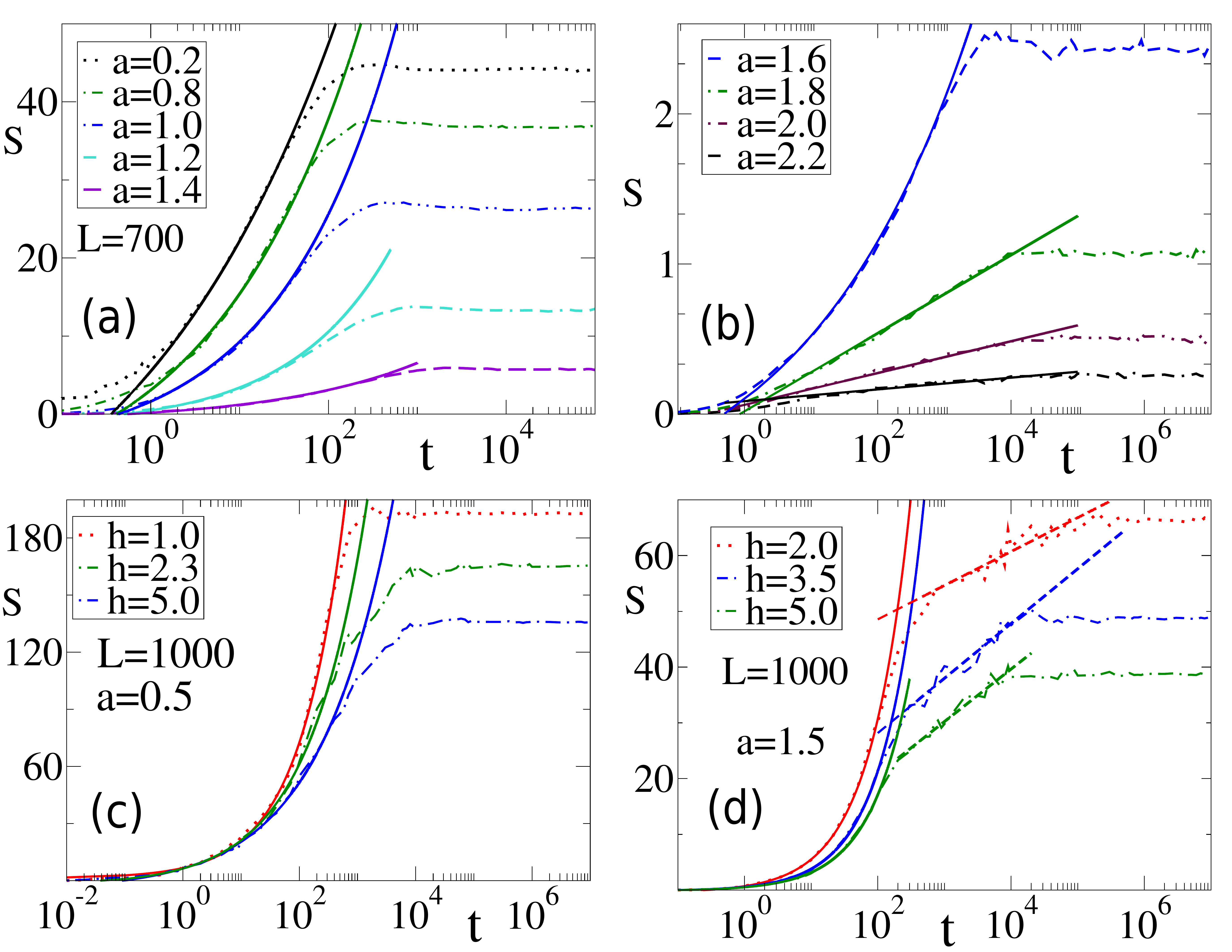}
\caption{(Color online) Time evolution of entanglement entropy $S(t)$  
for Model I is shown for  $0.2\le a \le 1.4$ in (a) and for $1.6 \le a \le 2.2$ in (b). Solid (dashed) lines  correspond 
to power-law $t^{\alpha}$-fit ($\log(t)$-fit).
Same results are shown for Model II in (c) with $a=0.5$ and in (d)
with $a=1.5$.
}
\label{fig1}
\end{figure}

\section{Entanglement entropy(EE)} \label{secIII}
In this section we will discuss the eigenstate EE and also the non-equilibrium dynamics of EE after a global quench starting
from above product state.
We note that a typical measure of the entanglement in a quantum system is bipartite von Neumann entanglement entropy
$S$ defined as, $S=-\Tr_{A}[\rho_A\ln\rho_{A}],$
where $ \rho_{A}=\Tr_{B}|\Psi\rangle\langle\Psi| $ is the 
reduced density matrix of a sub-system $A$ after dividing the system into two \textcolor{black}{equal adjacent}  parts $A$ and $B$, \textcolor{black}{ both 
comprised of $L/2$ sites}.
 $|\Psi\rangle$ is many-body wave function of the composite system. 

{\it{Eigenstate EE}}:
\textcolor{black}{
For model I, we  notice that the  typical eigenstate EE \cite{lev1,lev2,lev3,lucas.2019} 
(for details, see Ref.~\cite{suppl}) marked by \textcolor{black}{lines}
 in Fig.~\ref{fig3} (a) shows the absence of data collapse in  $S/L$. 
However,  the data collapse appears when we replace $S/L$ by $S/L^{\gamma (a)}$
with $\gamma(a)<1$ as shown in Fig.~\ref{fig3}(c). Interestingly, $\gamma(a)$ exhibits the duality  around $a=1$:  
 \textcolor{black}{$\gamma(a)\simeq\gamma(2-a)$ as shown in Fig.~\ref{fig3}(d) ( see Ref.~\cite{suppl} for more details)}. 
 This duality is the consequence of 
the duality present in the spatial exponent associated with the algebraically decaying long tail of SPSs for this model in either side of
$a<1$ and $a>1$ ~\cite{deng2017many}.
 \textcolor{black}{Furthermore, the inset of Fig.~\ref{fig3}(d) suggests that  
the expoent $\gamma$ 
  follows a universal behavior with spatial exponent 
 $a$ as long as SPSs are algebraically  localized.} The sub-extensive $L^\gamma$ law
 can be naively understood from the spatial algebraic structure of SPSs.
The total probability of finding a particle at any site $\in B$ 
while its localization center is  at site $x_0 \in A$ becomes $p \sim \sum_{x\in B} |x-x_0|^{-2 a} \sim L^{1-2a} f(x_0,a,L)$ where 
$f$ is a non-linear function of $x_0, a, L$. This type of fractional scaling with $L$ is absent 
for exponentially localized SPSs where $p \sim \xi f(x_0,\xi,L)$ and $\xi$ being the localization length. 
 Moreover, in the many-body case, EE becomes a
complex function of $x_0$ as the different SPSs have different localization centres. 
The absence of the coherent length scale $\xi$ in 
algebraic localization   can lead to non-trivial sub-extensive behavior in many-body EE. 
}

\textcolor{black}{On the other hand,   Fig. ~\ref{fig3}(b) 
 shows the presence of data collapse in  $S/L$ for model II. There exists a fraction of
delocalized ergodic phase yielding $S$ extensive. 
Even though, in both MF ($a<1$) and ME phase ($a>1$), $S\sim L$ [as shown in Fig.~\ref{fig3} (b)], 
one can distinguish them by their corresponding
 numerical values  of $S/L$. This is higher in MF phase
compared to ME phase  because all SPSs are essentially delocalized in MF phase.
Moreover, in both sides of 
$a=1$, transition between different $P_s$ phases are clearly visible.
The volume law of eigenstate  EE  has also been
recently observed for 1D short-range noninteracting model in the presence of correlated disorder, where
there exists a mobility edge in single particle spectrum \cite{soumi.18,deng2017many}.}


{\it {Asymptopic EE:}}
{We shall now extensively investigate the scaling of asymptotic saturation value  $S(L,t\to\infty)\equiv S_{\infty}$ 
 with $a$ as marked by \textcolor{black}{symbols}, starting from an initial product state $|\Psi_0\rangle$.
Figure ~\ref{fig3}(b) and (a) 
 show the presence of data collapse in  $S_{\infty}/L$  for model II and  absence of it in model I, respectively, 
 for different values of $L$. 
 Similar to the typical eigenstates, we here in dynamics find sub-extensive (extensive) nature of EE and duality
 in $\gamma$ for model I (model II) as shown in Fig.~\ref{fig3}(c) and (d) (Fig.~\ref{fig3}(b)). However, the proportionality
 factor associated with $S/L$ changes from 
 its eigenstate value. 
Since, these are noninteracting systems, we expect that $S_{\infty}$ obtained from dynamics
should show similar behavior as typical eigenstates. Note that
interaction leads to dephasing mechanism via scattering in the system and hence, for interacting system the above 
expectation may not hold true. MBL systems are one such examples, 
where we see that EE of many-body eigenstates obey area law, however, $S_{\infty}\sim L$\cite{serbyn.2013,bardarson.2012}.}
\textcolor{black}{Since, for both  models,
we do not have a parameter regime where all states are ergodic (delocalized), hence, $S_{\infty}/L$ 
and also  the  eigenstates EE density always become less than the \textcolor{black}{Page value~\cite{page.93}.}  
The EE associated with the mid-spectrum eigenstates of a generic interacting non-integrable systems obeying ETH,  
\cite{bera.2015,d2016quantum,rigol2012alternatives} satisfy this bound.}

{\it{Finite time rise:}}
\textcolor{black}{Having studied finite size scaling of aymptotic EE $S_{\infty}$, now we analyze the finite time growth of EE. The 
results are  shown for model I  in Fig.~\ref{fig1}(a), and (b).}
We observe for $a\le 1$, a power law rise occurs, $S \sim t^{\alpha}$. 
\textcolor{black}{This growth exponent $\alpha$ becomes larger near the point $a=1$ (see Ref.~\cite{suppl} for details)}. 
For $a>1$, 
growth exponent $\alpha$ decreases and for $a\sim 2$, EE shows a logarithmic rise. 
Note that in  the case of Hamiltonian \eqref{hamiltonian} even without disorder, initial growth of $S(t)$ is sub-linear
in $t$~\cite{essler.2016,suppl}.
Since, SPSs behave differently in either side of $a$; the 
presence of rapid fall of single particle wave function near the localization center causes 
relatively slow rise of EE for $1<a<2$ compared to $a<1$ regime~\cite{singh2017effect}. \textcolor{black}{We note that our power law 
growth of EE resembles  with  the out of time ordered correlator 
showing a deviation from light-cone like behavior 
in the context of long range models \cite{Luitz2019}.}

Turning into model II as shown in Fig.~\ref{fig1}(c), and (d), one can see a fast  
power law rise $(S\sim t^{\alpha_1})$ with $\alpha_1 <1$ followed by a much slower rise  in EE. 
We observe the value of $\alpha_1$ is larger for $a>1$ in comparison to  $a<1$ case.
We also note that in the MF phase, the growth exponent depends on $P_s$ phases,
 where as, in the ME phase, $\alpha$ remains almost same in different $P_s$ phases \textcolor{black}{(see Ref.~\cite{suppl} for details)}. 
 This is presumably the consequence of 
 the  fact that the spatial  structure of the mutifractal  SPS  are different in different $P_s$ phases. Contrastingly, for $a>1$ even in
 different $P_s$ ME phases,  the spatial structure of  localized states wave function are
 different from $a<1$.  For $a>1$, the long
 time slow growth,
 visible in a reasonably large time window, is found to be logarithmic.
 On the other hand, for $a<1$, we see a similar secondary slow 
rise of $S$, since the time window is much smaller
 we can not comment on it whether this is a power-law with exponent $\alpha_2 <\alpha_1$ or logarithmic.
One can naively connect our results with non-interacting central site  model \cite{central.2017}, where 
multifractality, appeared due to the coupling of a single central bound state with all  Anderson localized states, can give rise 
to a  slow logarithmic growth in entanglement dynamics.
Previously, logarithmic growth of $S$ was thought 
to be a unique feature of MBL systems\cite{bardarson.2012,prosen.2008}. Interestingly, our results indicate that
 the presence of two different types of SPS, there exists  a secondary slow rise 
  in the finite time evolution of $S$
 for model II; this growth is completely  missing for the uncorrelated disordered model I.

\textcolor{black}{Based on our anaysis of finite time EE in Model I and II, we can convey that the 
sub-linear temporal growth $S\sim t^{\alpha(\alpha_1)}$ is  related to the detail (structure and fraction of delocalized states) of 
SPSs in these systems: 1) This is apparently evident form the behavior of $S$ for $a< 3/2$ in Model I, and
2) for model II,
$\alpha_1$ remains unaltered with $a$ as long as one stays inside a fixed
 $P_s$ phase.  
However, the value of $\alpha$  interestingly changes as one varies $a$; similarly, $\alpha_1$ changes as one
goes from MF side to ME side even within the same $P_s$ phase.  
 A fnite fraction of delocalized SPSs can also cause the 
two-stage growth of EE in model II. On the other hand, this fraction becomes vanishly small for model I 
originating the single stage growth (see  Ref.~\cite{suppl} for detailed analysis). 
}


%

\begin{figure}
\includegraphics[width=0.5\textwidth,height=0.36\textwidth]{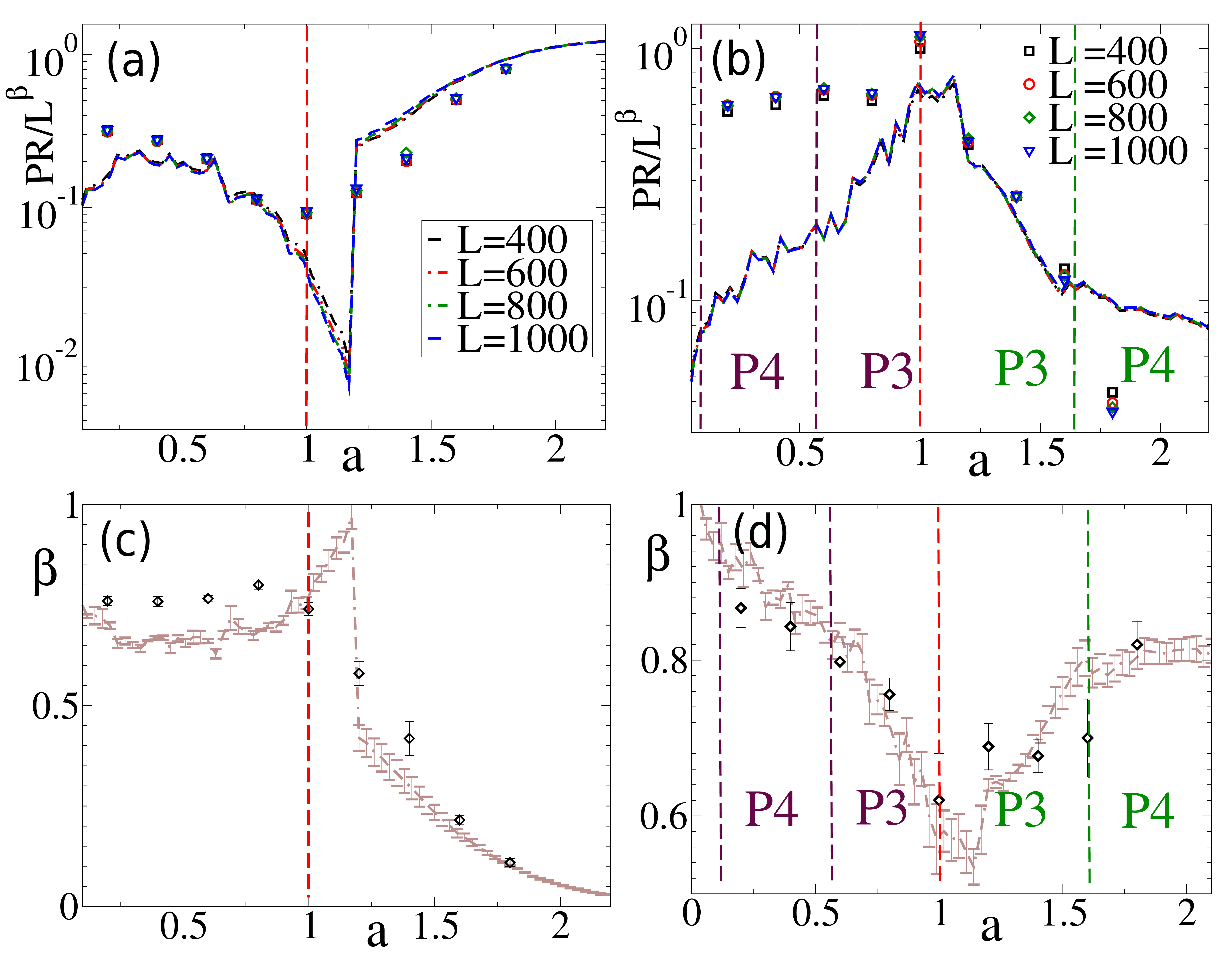}
\caption{(Color online)
 The sub-extensive scaling of PR is shown by the  data collapse of  PR/$L^{\beta}$ vs $a$ for  model I in (a)  
and  model II in (b). The   
variation of $\beta (a)$ with  $a$  for model I and II are depicted in (c) and (d), respectively. 
Lines (symbols) correspond to  PR obtained from eigenstates (dynamics). \textcolor{black}{$h=4.0$ is considered for Model II.}
 }
\label{fig4}
\end{figure}

\section{Participation ratio (PR)} \label{secIV}

Having examined the  scaling of eigenstate EE and  $S_{\infty}$  with $L$,
in the similar spirit,
we now look for
the scaling of eigenstate PR and the saturation values of PR (designated by PR$_{\infty}$).
We shall use the definition of many-body PR as introduced in Ref. \cite{bera.2015}.
It is defined for half-filling case as, PR$=L\bigl [2\sum_{\alpha=1}^{L}n_{\alpha}\sum_{j=1}^{L}|\phi_{\alpha}(j)|^4\bigr ]^{-1},$
where, $|\phi_{\alpha}\rangle$  and $n_{\alpha}$  are eigenvectors and eigenvalues  of one-body density matrix 
$\rho_{ij}=\langle \hat{c}^{\dag}_i\hat{c}^{}_{j}\rangle$ respectively. 
PR$\sim L$ for delocalized ergodic systems and PR$\sim\xi$  
for exponentially localized many-body states~\cite{celardo.2016,pr.2003}.




We study the characteristics of PR$_{\infty}$ and eigenstate PR 
in Fig.~\ref{fig4}.  From the data collapse of
PR/$L^{\beta}$ as a function of $a$ with different  system sizes (see Fig.~\ref{fig4} (a) and (b)),
we show for  both the models that PR exhibits a sub-extensive  scaling with $L$ \textcolor{black}{(see Ref.~\cite{suppl} for detail)}.
 In order to analyze the exponent $\beta$ more concretely, we show the variation of $\beta$ with $a$ for model I and II 
in Fig.~\ref{fig4} (c) and (d), respectively. 
$\beta$ remains fixed at a higher value for $a<1$ while it decreases monotonically for $a>1$ for model I. Very surprisingly,
unlike the EE, PR does not exhibit any duality with $a$ around $a=1$. 
The reason being PR is a local quantity, it is not able to capture the duality of SPSs in the  long-distance scale where
power law tail is observed in either side of $a=1$. Precisely,
PR accounts for the short distance behavior of SPSs where exponential and algebraic decay are present  for $a>1$  and $a<1$, respectively,
hence, $\beta$ is completely asymmetric 
around $a=1$.

On the other hand, for model II, $\beta$ shows a kind of symmetric behavior 
around $a=1$ (see Fig.~\ref{fig4}). This result may be counter intuitive in the sense that  phases in both sides of $a=1$ are 
completely different i.e., MF phase  for $a<1$ and ME phase  for $a>1$. 
Multifractal SPSs in this model have a form of  multiple sharp peak on the top of almost flat background in contrast to the ergodic
delocalized SPSs that are extended 
 all over the lattice \cite{suppl}. The structure of SPSs for MF phase is kind of similar to the exponentially localized SPSs having 
 only one peak and the background is suppressed exponentially with distance from that peak. 
Since, PR is an inappropriate measure to identify the fine tuned long distance structure of SPSs for ME and MF phases,
we find similar behavior of $\beta$ in either side of $a$. However, we note that $\beta$ is much
closer to $1$ in MF phase  compared to ME phase.  Moreover, from the 
variation of $\beta$ with $a$,
we can roughly identify different $P_s$ phases in either side of $a=1$ [see Fig.~\ref{fig4}(d)].
\textcolor{black}{Moreover, we note that in the calculation with typical eigenstates, we
discard  a few bottom  spectrum delocalized states to minimize their effect 
 (see Ref.~\cite{suppl} for details). On the other hand, all the energy states come automatically into
the non-eignestate dynamics.  We believe that this is the origin of the apparent dissimilarities  between the predictions from typical
eigenstate and long time dynamics as observed in Fig.~\ref{fig3} and Fig.~\ref{fig4}.
}

\begin{table}[t]
 \begin{tabular}{ c |c|c}
 SPS & $\qquad S_{\infty} \qquad$ & $\qquad PR_{\infty} \qquad  $\\
  \hline            
  Exponential localization  & $L^0$ & $L^0$  \\
   Algebraic localizaion & $L^{\gamma}$, $\gamma<1$ & $L^{\beta} $, $\beta<1$ \\
  Ergodic (delocalized) & $L$ & $L$ \\
 Multi-fractal (non-ergodic) & $L$ & $L^{\beta}$, $0<\beta\le 1$  \\
 Mobility-edge & $L$ & $L^{\beta}$, $0<\beta\le 1$  \\
  \hline
 \end{tabular}
\caption{Summary of the main differences 
between different phases in the non-interacting long range systems. We note that $\beta$
and $\gamma$ are model dependent exponents.}
\label{tb}
\end{table}

\paragraph*{Conclusion:}
We summarize our main results in Table.~\ref{tb}.
One of the most intriguing finding  is to show the sub-extensive  scaling in
EE and PR, when SPSs are algebraically localized  as observed in model I.
This is firmly evident from both the eigenstate and long time dynamics. 
\textcolor{black}{The absence of length scale thus
 imprints its'  signature  unlike the exponentially localized phase.
 Moreover, these behaviors are not the artifact of delocalized states present in Model I (at least for $a<3/2$) as  the number of such states has
measure  zero  for $L\to \infty$ \cite{nol1,Ossipov_2013,deng.2018}.}
\textcolor{black}{ Turning to model II, asymptotic and eigenstate  EE both obey volume law due to the presence of ergodic SPSs; 
however,  interestingly, the proportionality factors change in
 different $P_s$ phases.
The adiabatic connectivity allows us to conjecture that algebraically localized quasi-local integrals of motion would survive 
even in the weakly interacting limit \cite{serbyn.2013,Imbrie2016,modak2016integrals,de2019algebraic}
 and hence,  the eigenstate EE scaling should remain unaltered even in the above limit. 
One might not expect the similar scaling of asymptotic EE,
obtained from the long time dynamics, due to the dephasing mechanism caused by the interaction.
 }

Our study further reveals the connection between the  exponents ($\gamma$ for EE and 
$\beta$ for PR), and the spatial structure of the SPSs. The
EE is maximally governed by the long distance nature of the SPSs and thus the duality in $\gamma$ is 
closely connected to the duality of the localization exponent $\nu$ as noticed for model I \cite{deng.2018}.
Moreover, $\gamma$ follows a universal behavior as far as the algebraically localized SPSs are concerned.
In contrary, PR  captures the short distance nature of  
correlation  \textcolor{black}{leading to the fact that 
exponent $\beta$ does not show duality around $a=1$.  The
short distance behavior of SPSs  are very different for $a<1$
and $a>1$ for model I.  Surprisingly, model II shows  duality like behavior
within a small window around
$a= 1$. This can be related to the peculiar
spatial distribution of multifractal SPSs at short distance (see Ref.~\cite{suppl} for detail).}

\textcolor{black}{ Another important contribution of our work is to show how 
the structure of SPSs can influence the finite time rise of EE. An unprecedented    two-stage 
growth of EE for model II is exclusively observed while model I exhibits single-stage growth.
The secondary rise in  EE for model II might be related to the fact that there exist finite fraction of two 
types of SPSs i.e., multifractal and delocalized or localized and delocalized.
The initial algebraic temporal growth is common in both  the models. Recent studies also  find  signatures of temporal 
 power law growth of EE in long range interacting  models \cite{naini19,de2019algebraic}. 
The connection between the temporal power-law  growth of EE and algebraical SPSs (LIOM) for non-interacting (interacting) model 
is still an open field of research. Given the experimental realizability of spin models \cite{exp5,exp6,exp7,exp8}, 
we believe that our study would initiate a plethora of work
in this direction. }



\section{Acknowledgements}  
Authors  thank P. Calabrese and  L. Vidmar for reading
the manuscript and for several comments.  The authors also thank the anonymous referee for useful comments. 

\bibliography{reference}

\end{document}